\documentclass[12pt, 
a4paper]{article}

\usepackage{amsmath, amssymb, amsthm}
\usepackage{graphicx}
\usepackage{hyperref}
\usepackage{float}
\usepackage{braket}
\usepackage{mathtools}
\usepackage{bm}
\usepackage{caption}
\usepackage{subcaption}
\usepackage{siunitx}
\usepackage{tcolorbox}
\usepackage{tikz-cd}
\usepackage[margin=1in]{geometry}
\usepackage{titlesec}
\titleformat{\section}
  {\normalfont\Large\bfseries\centering} % formatting: size, bold, centered
  {\thesection}{1em}{}                   % section number formatting

\usepackage{chngcntr}
\renewcommand{\thesection}{\arabic{section}}

\usepackage{listings}
\usepackage{xcolor}

\definecolor{codegray}{gray}{0.95}
\definecolor{deepblue}{rgb}{0.0,0.0,0.5}
\definecolor{deepred}{rgb}{0.6,0.0,0.0}
\definecolor{deepgreen}{rgb}{0.0,0.5,0.0}

\usepackage[backend=biber]{biblatex}
\newcommand{\hcut}{\hbar}
\newcommand{\sigmaX}{\sigma_x}

\newcommand{\proj}[1]{\frac{1 \pm \sigmaX^{#1}}{2}} % Projector notation

\title{\LARGE \textbf{1D Cluster State Generation On Superconducting Hardware}}
\author{Rahul Dev Sharma\thanks{Department of Physics, Shahjalal University of Science and Technology, Sylhet, Bangladesh; Email: rahul.ds.science@gmail.com}
}

\date{}
\begin{document}

\maketitle

    % %% Author and Supervisor
    % \begin{minipage}[t]{0.45\textwidth}
    %     \centering
    %     \textbf{Submitted by} \\[0.2cm]
    %     \textbf{Rahul Dev Sharma} \\
    %     Reg. No.: 2019132048 \\
    % \end{minipage}
    % \hfill

    % \vfill

% \end{titlepage}

%% Acknowledgements Section
\pagenumbering{arabic} % restore page numbers
% \begin{center}
% \LARGE\textbf{Acknowledgements}
% \end{center}
% \noindent I would like to express my heartfelt gratitude to all those who have contributed to the successful completion of this undergraduate research project.

% First and foremost, I extend my sincere appreciation to my supervisor, \textbf{Dr. Jaseer Ahmed}, for their invaluable guidance, continuous support, and constructive feedback throughout this research endeavor.

% I am deeply grateful to \textbf{Md. Sakibul Islam Sazzad}, PhD candidate at \textbf{University of Central Florida, USA}, who served as my external mentor. Their intellectual insights, research expertise, and willingness to share knowledge significantly enriched my understanding of the subject matter and contributed to the quality of this project.

% My sincere appreciation goes to my fellow students and friends, particularly \textbf{Sakibul Islam Rayhan, EEE, SUST}, for their unwavering moral support and encouragement throughout this challenging yet rewarding journey. Their presence made the difficult moments more bearable and the achievements more meaningful.

% Finally, I express my gratitude to my family for their patience, understanding, and constant encouragement, which provided me with the strength to persevere through this academic pursuit.

\begin{abstract}
Measurement-based Quantum Computation(MBQC) utilize entanglement as resource for performing quantum computation. Generating cluster state using entanglement as resource is a key bottleneck for the adoption of MBQC. To generate cluster state with charge-qubit arrrays, we provide analytical  derivations and numerical validations for 4-qubit cluster state. We compare our fidelities under ideal (noise-free) Hamiltonian evolution and due to effect of decoherence. We show incorporating energy relaxation ($T_1$) yields $>$90\% fidelity while pure dephasing \(T_2\) show \(70\%\) decays at fourth harmonics.  We further show under noise \(T_2\) decays to 50\% within 15 time units, versus $>$70\% under relaxation time units (\(T_1\))--only. This decay quantify degradation effect of $T_2$ on preparing cluster--state preparation is more than $T_1$. We highlight the critical need for targeted error-mitigation strategies in near-term MBQC implementations.
\end{abstract}

% \tableofcontents

% Introduction section
\section{Introduction}
Quantum computing (QC) harnesses quantum mechanics to achieve computational capabilities far beyond classical computers. Unlike classical bits, restricted to states of 0 or 1, quantum bits (qubits) exploit superposition, enabling simultaneous processing of multiple states. Combined with entanglement and interference, this allows quantum computers to address complex problems, such as factoring large numbers \cite{shor1997} and simulating quantum systems \cite{feynman1982}, exponentially faster than classical counterparts \cite{Nielsen_Chuang_2010}. However, qubits are susceptible to decoherence, necessitating robust computational models \cite{preskill2018quantum}.

There are various formalism of Quantum Computation. Such as, gate model of quantum computation, adiabatic quantum computation\cite{RevModPhys.90.015002}, measurement-based quantum computation (MBQC), topological quantum computation\cite{freedman2002topologicalquantumcomputation}, and quantum walks\cite{Venegas_Andraca_2012}. In terms of computational equivalence and implementation feasibility, gate model and measurement based model are the best two. Though gate based approach is widely adopted, it has some issues. High gate overhead, makes it easier for errors to enter into the computation. Whereas in measurement based quantum computation, the overhead is drastically reduced, thus solving the gate overhead problem.

Measurement-based quantum computing (MBQC), or one-way quantum computing, introduced by Raussendorf and Briegel \cite{Raussendorf2001}, offers a promising alternative to the gate-based model. In MBQC, computation is driven by single-qubit measurements on a pre-prepared entangled resource state, typically a cluster state. Cluster states, graph-like entangled structures where nodes represent qubits and edges denote entangling operations, serve as universal resources for MBQC \cite{Raussendorf2001}. By selecting appropriate measurement bases, any quantum computation can be executed \cite{briegel2009measurement}.

MBQC is advantageous in systems where two-qubit gates are challenging but single-qubit measurements are reliable. Superconducting quantum circuits, a leading QC platform, face difficulties with high-fidelity two-qubit gates, making MBQC an attractive approach \cite{devoret2013superconducting}. By preparing cluster states offline, MBQC simplifies quantum algorithm implementation in such systems \cite{briegel2009measurement}. The model’s versatility is evident in its exploration across photonic\cite{Li2025}, trapped-ion \cite{lanyon2008measurement} \cite{tame2008measurement}, and neutral atom platforms \cite{Graham2024}.

You et al. proposed an efficient one-step method for generating large cluster states using superconducting quantum circuits \cite{PhysRevA.75.052319}. By briefly activating interqubit coupling, their approach creates cluster states robust against parameter variations, ideal for solid-state quantum computing \cite{devoret2013superconducting}. Given the critical role of cluster states in MBQC and the prominence of superconducting circuits, revisiting this work is timely.

This paper recreates You et al.’s study, enhancing it with detailed derivations, equation verification, and simulations assessing decoherence effects (T1 and T2) \cite{PhysRevA.75.052319, devoret2013superconducting}. The paper is structured as follows: Section 2 outlines the theoretical background, Section 3 presents derivations, Section 4 details verification, Section 5 discusses simulation results, and Section 6 concludes with future directions.

\section{Theoretical Background}

\subsection{Fundamentals of Quantum Computing}

Quantum computing leverages the principles of quantum mechanics to perform computations that are intractable for classical computers. The fundamental unit of quantum information is the quantum bit, or qubit, which differs from classical bits in its ability to exist in superposition.

\subsubsection{Qubits and Quantum States}

A single qubit is represented by a state vector in a two-dimensional complex Hilbert space. The computational basis states \(\ket{0}\) and \(\ket{1}\) correspond to the classical bit values; however, a general qubit state is a linear combination
\[
\ket{\psi} = \alpha \ket{0} + \beta \ket{1},
\]
with complex amplitudes \(\alpha\) and \(\beta\) satisfying \(\lvert\alpha\rvert^2 + \lvert\beta\rvert^2 = 1\). This superposition principle enables a qubit to encode both classical states simultaneously until measurement. In an \(n\)-qubit register, the Hilbert space dimension grows as \(2^n\), allowing an \(n\)-qubit system to occupy
\[
\ket{\psi} = \sum_{i=0}^{2^n - 1} \alpha_i \ket{i},
\]
where \(\sum_i \lvert\alpha_i\rvert^2 = 1\).

\subsubsection{Quantum Entanglement}

Entanglement is a uniquely quantum correlation with no classical counterpart. In entangled states, the global state cannot be factorized into individual qubit states. For instance, the Bell state
\[
\ket{\Phi^+} = \frac{1}{\sqrt{2}}\bigl(\ket{00} + \ket{11}\bigr)
\]
exhibits perfect correlations: a measurement on one qubit instantaneously determines the outcome of the other, regardless of spatial separation. Entanglement is a critical resource for quantum algorithms, underpinning speedups over classical computation.

\subsubsection{Quantum Measurement}

Measurement projects a qubit state onto an eigenbasis, collapsing superposition with probabilities given by the squared amplitudes. In the computational basis, \(\alpha\ket{0} + \beta\ket{1}\) yields \(\ket{0}\) with probability \(\lvert\alpha\rvert^2\) and \(\ket{1}\) with probability \(\lvert\beta\rvert^2\). Alternative bases, such as the diagonal basis \(\{\ket{+},\ket{-}\}\) where \(\ket{\pm} = (\ket{0}\pm\ket{1})/\sqrt{2}\), are essential in certain computational protocols. 

\subsubsection{Quantum Gates and Circuits}

Quantum algorithms are implemented via unitary transformations on qubit states. Common single-qubit operations include the Pauli matrices
\[
X = \begin{pmatrix}0 & 1\\1 & 0\end{pmatrix},\quad
Y = \begin{pmatrix}0 & -i\\i & 0\end{pmatrix},\quad
Z = \begin{pmatrix}1 & 0\\0 & -1\end{pmatrix},
\]
together with the Hadamard \(H\), phase \(S\), and \(\pi/8\) \(T\) gates:
\[
H = \frac{1}{\sqrt{2}}\begin{pmatrix}1 & 1\\1 & -1\end{pmatrix},\quad
S = \begin{pmatrix}1 & 0\\0 & i\end{pmatrix},\quad
T = \begin{pmatrix}1 & 0\\0 & e^{i\pi/4}\end{pmatrix}.
\]
Continuous rotations \(R_x(\theta)\), \(R_y(\theta)\), and \(R_z(\theta)\) are generated by the corresponding Pauli operators. Two-qubit interactions, such as the controlled-NOT
\[
\mathrm{CNOT} = \begin{pmatrix}
1 & 0 & 0 & 0\\
0 & 1 & 0 & 0\\
0 & 0 & 0 & 1\\
0 & 0 & 1 & 0
\end{pmatrix}
\]
and the controlled-\(Z\)
\(\mathrm{CZ} = \mathrm{diag}(1,1,1,-1)\), create entanglement. The combination of arbitrary single-qubit rotations and an entangling two-qubit gate constitutes a universal gate set.

\subsection{Measurement-Based Quantum Computing (MBQC)}

Measurement-based quantum computing (MBQC), or one-way quantum computing, employs a pre-prepared, highly entangled multi‑qubit resource state to drive all computational steps via single‑qubit measurements \cite{Raussendorf2001, briegel2009measurement}. In contrast to the circuit model—which implements unitary gates such as Hadamard and CNOT—MBQC consumes qubits by projective measurement, and the remaining entanglement propagates quantum information.

The canonical resource is a cluster state, whose entanglement structure underpins universality. Qubits are measured sequentially: once a qubit is measured, it is removed from the resource and yields a classical outcome. Crucially, each measurement basis depends on prior outcomes, implementing an adaptive feed‑forward that guarantees deterministic execution of the intended algorithm.

A measurement in the Pauli \(X\) basis, with eigenstates
\[
\ket{+} = \frac{1}{\sqrt{2}}\bigl(\ket{0} + \ket{1}\bigr), 
\quad
\ket{-} = \frac{1}{\sqrt{2}}\bigl(\ket{0} - \ket{1}\bigr),
\]
can simulate a Hadamard gate by projecting the cluster state accordingly \cite{Raussendorf2001}. More general single‑qubit rotations arise from measurements in bases in the \(XY\) plane, parameterized by an angle \(\theta\).

MBQC proves advantageous in settings where multi-qubit gates are difficult to implement with high fidelity, while single-qubit measurements can be performed reliably. With only three essential steps—state preparation, entanglement, and measurement—it offers a streamlined and scalable model of quantum computation \cite{Raussendorf2001}.

\subsection{Cluster States: Definition and Properties}

Cluster states are a foundational resource in quantum information science, particularly for measurement-based quantum computing (MBQC)~\cite{Raussendorf2001, briegel2009measurement}. These highly entangled multi-qubit states are defined on a graph \( G=(V,E) \), where each vertex \( v \in V \) represents a qubit and each edge \( e \in E \) signifies an entangling operation. Typically, this entangling operation is a controlled-Z (CZ) gate, defined as

\[
\text{CZ}_{i,j} = \ket{0}\bra{0}_i \otimes I_j + \ket{1}\bra{1}_i \otimes Z_j.
\]

The CZ gate applies a phase of \( -1 \) to the \( \ket{11} \) component of a two-qubit state, thereby generating entanglement~\cite{Nielsen_Chuang_2010}.

\subsubsection{Formation of Linear Cluster States}

For a linear cluster state with \( N \) qubits, the underlying graph is a simple chain, where each qubit is connected only to its nearest neighbors. The state is prepared by first initializing each qubit in the superposition state

\[
\ket{+} = \frac{1}{\sqrt{2}}(\ket{0} + \ket{1}),
\]

which is an eigenstate of the Pauli \( X \) operator, i.e., \( X\ket{+} = \ket{+} \). Subsequently, CZ gates are applied between all adjacent qubits along the chain:

\[
\ket{\text{C}_N} = \prod_{i=1}^{N-1} \text{CZ}_{i,i+1} \left( \bigotimes_{i=1}^N \ket{+}_i \right).
\]

In the computational basis, this state can be explicitly expressed as

\[
\ket{\text{C}_N} = \frac{1}{2^{N/2}} \sum_{x \in \{0,1\}^N} (-1)^{\sum_{i=1}^{N-1} x_i x_{i+1}} \ket{x_1 x_2 \dots x_N},
\]

where the phase factor \( (-1)^{\sum_{i=1}^{N-1} x_i x_{i+1}} \) introduces correlations between neighboring qubits. This phase structure encodes the intricate entanglement essential for MBQC~\cite{Raussendorf2001}.

\vspace{0.5em}
% \textbf{[Insert Figure 1: Schematic representation of a linear cluster state with N qubits, showing qubits as vertices connected by CZ gates.]}

\subsubsection{Stabilizer Formalism}

Cluster states are a prime example of \emph{stabilizer states}, meaning they are unique eigenstates with eigenvalue \( +1 \) of a set of commuting Pauli operators known as stabilizers~\cite{Gottesman1997}. For a linear cluster state, the stabilizer operators provide a robust method for characterizing and verifying the state's properties.

For an interior qubit \( i \) such that \( 1 < i < N \), the stabilizer operator is

\[
S_i = Z_{i-1} \otimes X_i \otimes Z_{i+1}.
\]

The linear cluster state \( \ket{\text{C}_N} \) satisfies \( S_i \ket{\text{C}_N} = \ket{\text{C}_N} \) for all \( i \). For the boundary qubits, the stabilizers are modified to reflect their connectivity. For the first qubit \( i=1 \), the stabilizer is

\[
S_1 = X_1 \otimes Z_2.
\]

For the last qubit \( i=N \), the stabilizer is

\[
S_N = Z_{N-1} \otimes X_N.
\]

This stabilizer formalism is crucial not only for defining cluster states but also for their experimental verification and for understanding their robustness against certain types of errors~\cite{VanDenNest2007}.

\subsection{Superconducting Quantum Circuits}
Superconducting quantum circuits are a prominent platform for quantum computing, leveraging the unique properties of superconductors to create and manipulate quantum bits (qubits). These circuits operate at cryogenic temperatures, typically around 20 mK, where materials such as aluminum and niobium exhibit superconductivity, characterized by zero electrical resistance and the expulsion of magnetic fields via the Meissner effect \cite{devoret2013superconducting}. This subsection introduces superconducting qubits, their types, and their relevance to quantum computing, with a focus on their application in generating cluster states for measurement-based quantum computing (MBQC).

\subsubsection{Introduction to Superconducting Qubits}

Superconducting qubits are artificial two-level quantum systems engineered from superconducting circuits, designed to have discrete energy levels that serve as the computational basis states \(|0\rangle\) and \(|1\rangle\). Their scalability, compatibility with semiconductor fabrication techniques, and relatively long coherence times make them a leading choice for quantum computing applications \cite{Clarke2008}. Superconducting qubits exploit the Josephson effect, where Cooper pairs tunnel through a thin insulating barrier, providing the nonlinearity necessary for quantum operations.

The most common types of superconducting qubits include:

\begin{itemize}
    \item \textbf{Transmon Qubits}: These qubits are designed to be insensitive to charge noise, offering coherence times of tens to hundreds of microseconds. They are widely used in modern quantum processors due to their robustness and ease of fabrication \cite{Koch2007}.
    \item \textbf{Flux Qubits}: These utilize the magnetic flux through a superconducting loop to define their states, providing high tunability but sensitivity to flux noise \cite{Clarke2008}.
    \item \textbf{Charge Qubits}: Also known as Cooper-pair boxes, these are based on the number of Cooper pairs on a superconducting island. They were foundational in early quantum computing research but are sensitive to charge noise \cite{Makhlin2001}.
    \item \textbf{Phase Qubits}: These rely on the phase difference across a Josephson junction, offering an alternative design but are less common in current systems \cite{Martinis2002}.
\end{itemize}

Our work focuses on superconducting charge qubits, which—despite their well-documented susceptibility to charge noise (later largely addressed by transmon designs\cite{Schreier2007SuppressingCN}), to deliver a comprehensive, pedagogically oriented derivation and simulation of the one-step cluster-state generation protocol.

\subsubsection{Superconducting Charge Qubits}
A superconducting charge qubit (often called a Cooper-pair box) is really just a tiny superconducting island that can hold an integer number of excess Cooper pairs, weakly connected to a large superconducting “reservoir” (or ground) through one or more Josephson junctions.

\paragraph{The Island and the Reservoir\newline}

\begin{itemize}
    
\item Island: A small piece of superconductor whose total charge (in units of Cooper pairs) is quantized. We call the two lowest charge states 
$|0\rangle$ (no extra pair) and 
$|1\rangle$ (one extra pair). These two states form the computational basis of the qubit.

\item Reservoir: A bulk superconducting electrode that acts as a charge “ground” or reference. Cooper pairs can tunnel back and forth between island and reservoir through the Josephson junction(s).

\end{itemize}

\section{Derivation \& Verification of Cluster State Generation via Hamiltonian Evolution}
\subsection{General Hamiltonian Formulation}

We start with the general Ising-like Hamiltonian for a chain of qubits, as presented in Ref.~\cite{PhysRevA.75.052319}:
\begin{subequations}
\begin{equation} \label{eq:1}
H = \hbar g(t) \sum_{i,j} \Gamma(i-j) \proj{i} \proj{j},
\end{equation}
where
\begin{itemize}
    \item \(\hbar\) is the reduced Planck's constant,
    \item \(g(t)\) is a time-dependent function that modulates the interaction strength,
    \item \(\Gamma(i-j)\) represents the spatial interaction profile,
    \item \(\proj{}\) denotes the projection operator~\cite{Nielsen_Chuang_2010}.
\end{itemize}

Due to anisotropy, where both the Ising interaction and the external magnetic field are aligned along the $x$-direction, the spins align in the $|+\rangle$ or $|-\rangle$ basis. Here,
\(|+\rangle = \frac{1}{\sqrt{2}} (|0\rangle + |1\rangle)\) and \(|-\rangle = \frac{1}{\sqrt{2}} (|0\rangle - |1\rangle)\).

The projection operator selects only the relevant eigenstates of \(\sigma_i^x\), filtering out the rest. Specifically,
\(\Pi_{+} = \frac{1 + \sigma_i^x}{2}\) and \(\Pi_{-} = \frac{1 - \sigma_i^x}{2}\), such that:
\begin{align*}
    \Pi_{+}|+\rangle &= |+\rangle, & \Pi_{+}|-\rangle &= 0, \\
    \Pi_{-}|+\rangle &= 0, & \Pi_{-}|-\rangle &= | - \rangle.
\end{align*}

The product \(\proj{i} \proj{j}\) in Eq.~\eqref{eq:1} considers all possible spin pair configurations: $|++\rangle$, $|+-\rangle$, $|-+\rangle$, and $|--\rangle$.

By expanding and simplifying Eq.~\eqref{eq:1}, we obtain:
\begin{align}
H = \hbar g(t) \sum_{i,j} \Gamma(i-j) \left[ \frac{1}{4} (1 \pm \sigma_i^x \pm \sigma_j^x + \sigma_i^x \sigma_j^x) \right].
\end{align}

The constant term $\frac{1}{4}$ contributes an overall energy shift and can typically be neglected in dynamic analyses. This leads to:
\begin{equation}\label{eq:1c}
H = \hbar g(t) \sum_{i,j} \Gamma(i-j) \left[ \pm \frac{1}{4} (\sigma_i^x + \sigma_j^x) + \frac{1}{4} \sigma_i^x \sigma_j^x \right].
\end{equation}

The first term accounts for single-qubit effects, such as those arising from control fields, while the second term captures Ising-type two-qubit interactions.

Restricting to nearest-neighbor coupling:
\begin{equation}\label{eq:1d}
\Gamma(i-j) = J_{i,j}\delta_{|i-j|,1},
\end{equation}
with $J_{i,j}$ representing the coupling strength between qubits $i$ and $j$.

Substituting Eq.~\eqref{eq:1d} into Eq.~\eqref{eq:1c}, we obtain:
\begin{equation}
H = \hbar g(t) \sum_{i,j} J_{i,j} \delta_{|i-j|,1} \left[ \pm \frac{1}{4}(\sigma_i^x + \sigma_j^x) + \frac{1}{4} \sigma_i^x \sigma_j^x \right].
\end{equation}

This expression simplifies further to:
\begin{equation}
H = \hbar g(t) \sum_{i} J_{i,i+1} \left[ \pm \frac{1}{4}(\sigma_i^x + \sigma_{i+1}^x) + \frac{1}{4} \sigma_i^x \sigma_{i+1}^x \right],
\end{equation}
noting that both $i+1$ and $i-1$ contribute similarly, and focusing on $i+1$ avoids redundancy.
\end{subequations}

Combining single-qubit terms, we write:
\begin{subequations}
\begin{equation}
H_i = \pm \frac{1}{4} \hbar g(t) \sigma_i^x,
\end{equation}

defining an effective interaction coefficient:
\begin{equation}
\Lambda_{i,i+1} = \frac{1}{4} \hbar g(t) J_{i,i+1}.
\end{equation}

The final Hamiltonian becomes:
\begin{equation}
H = \sum_{i=1}^N \left[ H_i + \Lambda_{i,i+1} \sigma_i^x \sigma_{i+1}^x \right],
\end{equation}
with $\Lambda_{N,N+1} = 0$ since there is no $(N+1)$-th spin.

\end{subequations}

Now the Hamiltonian of the $i$th Charge Qubit can be written as,

\begin{equation}
H_i = \varepsilon_i(V_i) \sigma_z^{(i)} - \bar{E}_{J_i} \sigma_x^{(i)}
\end{equation}

where,
\begin{equation}
\begin{aligned}
\varepsilon_i(V_i) &= \tfrac{1}{2} E_{c_i} \Bigl(\tfrac{C_i V_i}{e} - 1\Bigr), \\
\bar{E}_{J_i}      &= E_{J_i} \cos\!\Bigl(\tfrac{\pi \Phi_i}{\Phi_0}\Bigr)
\end{aligned}
\end{equation}

Here, $E_{c_i}$ is the charging energy(energy needed to put extra cooper pair from the reservoir into the superconducting islands) of the superconducting island in the $i$th qubit, and $E_{J_i}$ is the Josephson coupling energy(it is a measure of the
difficulty for Cooper pairs to tunnel across the junction) of the two identical Josephson junctions connected to the island. $V_i$ is the biasing gate voltage applied to the qubit, One of the capacitances, \( C_i \), connects the island to an external voltage source. This creates a continuous offset charge given by
$
\frac{C_i V_i}{e},
$
 and $\Phi_i$ is the externally applied magnetic flux through the small loop associated with the $i$th qubit, which modulates the value of $E_{J_i}$.

When a Cooper pair tunnels onto or off the island through the Josephson junction, it changes the number of excess Cooper pairs by one. In the two-level picture:

\begin{itemize}
  \item \textbf{Tunneling onto the island:}
  \[
  |0\rangle \longrightarrow |1\rangle
  \]
  \item \textbf{Tunneling off the island:}
  \[
  |1\rangle \longrightarrow |0\rangle
  \]
\end{itemize}

This tunneling is induced by the Josephson coupling energy \( E_J \). In the Hamiltonian, this process is described by the operator \( \sigma^x \), which acts as:
\[
\sigma^x = |0\rangle\langle1| + |1\rangle\langle0|
\]
This operator causes coherent transitions (flipping) between the \( |0\rangle \) and \( |1\rangle \) states, effectively representing the quantum tunneling of Cooper pairs.

We assume that the charge qubit works in the charging regime with $E_{ci}\gg
E_{Ji}$.

Flux-Dependent Interqubit Coupling is,
\begin{equation}
\Lambda_{i,i+1} = L_J \left( \frac{\pi}{\Phi_0} \right)^2 E_{J_i} E_{J_{i+1}} \sin\left( \frac{\pi \Phi_i}{\Phi_0} \right) \sin\left( \frac{\pi \Phi_{i+1}}{\Phi_0} \right)
\end{equation}

Here, the large Josephson junction acts as an effective inductance
\[
L_J = \frac{\Phi_0}{2\pi I_0}, \quad \text{with} \quad I_0 = \frac{2\pi E_{J_0}}{\Phi_0}
\]
being the critical current of the large junction.

When each charge qubit is tuned to the degeneracy point $\left(C_i V_i / e = 1\right)$, the Hamiltonian of the array becomes:
\begin{equation}
H_A = \sum_{i=1}^{N} \left( -\bar{E}_{J_i} \sigma_x^{(i)} + \Lambda_{i,i+1} \sigma_x^{(i)} \sigma_x^{(i+1)} \right)
\end{equation}

Now we will impose a relationships between the Josephson energies (\(\overline{E}_{Ji}\)) and coupling strengths (\(\Lambda_{i,i+1}\)) to transform the Hamiltonian into a form suitable for cluster state generation. Lets tune the parameters so that,

\[
\frac{1}{2} \overline{E}_{Ji} = \Lambda_{i,i+1} \equiv \frac{1}{4} \hbar g, \quad i = 2, 3, \ldots, N-1
\]
\[
\overline{E}_{J1} = \Lambda_{1,2} = \overline{E}_{JN} \equiv \frac{1}{4} \hbar g \tag{7}
\]

The effective Josephson energy $\bar{E}_{J_i}$ decreases from its maximum $E_{J_i}$ down to zero as you thread flux $\Phi_i$ from $0$ to $\Phi_0/2$.

The coupling $\Lambda_{i,i+1}$ increases from zero up to its maximal inductive value
\[
\Lambda_{i,i+1}^{\text{max}} = \frac{\Phi_0^2}{(2\pi)^2 L_J} \cdot \frac{1}{E_{J_i} E_{J_{i+1}}}
\]
as you tune those same fluxes.

Because one function ($\bar{E}_{J_i}$) is monotonic decreasing in $\Phi_i$ and the other ($\Lambda_{i,i+1}$) is monotonic increasing in $\Phi_i$, there is always a unique flux value $\Phi_i \in [0, \Phi_0/2]$ at which
\[
\Lambda_{i,i+1}(\Phi_i) = \frac{1}{2} \bar{E}_{J_i}(\Phi_i).
\]

These conditions adjust the single-qubit and interaction terms in equation (6) to balance their contributions, particularly at the boundaries (qubits 1 and \( N \)). The condition \( \frac{1}{2} \overline{E}_{Ji} = \Lambda_{i,i+1} \) for internal qubits ensures that the single-qubit terms (\(-\overline{E}_{Ji} \sigma_i^x\)) and coupling terms (\(\Lambda_{i,i+1} \sigma_i^x \sigma_{i+1}^x\)) are related, allowing the Hamiltonian to be rewritten in a projector form. For boundary qubits, setting \(\overline{E}_{J1} = \Lambda_{1,2}\) and \(\overline{E}_{JN} = \Lambda_{N-1,N}\) adjusts the edge interactions, accounting for the lack of coupling beyond qubits 1 and \( N \).

% Equation 8 rewrites the Hamiltonian using the conditions from equation 7:

% \[
% H_A = \hbar g \sum_{i=1}^{N-1} \frac{1 - \sigma_i^x}{2} \frac{1 - \sigma_{i+1}^x}{2} \tag{8}
% \]

Now let's use the conditions stated above.

We have previously got equation 6,

\[
H_A = \sum_{i=1}^N  -\overline{E}_{Ji} \sigma_x^{(i)}  + \sum_{i=1}^{N-1} \Lambda_{i,i+1} \sigma_x^{(i)} \sigma_x^{(i+1)} \tag{6}
\]

Now, lets, replace the \(\overline{E}_{Ji}\), \(\Lambda_{i,i+1}\) values.

\[
H_A = -\frac{1}{4} \hcut g\sigma^{(1)}_x - \frac{1}{2} \hcut g\sum_{i=2}^{N-1}   \sigma_x^i -\frac{1}{4} \hcut g\sigma^{(N)}_x  + \frac{1}{4} \hcut g\sum_{i=1}^{N-1}  \sigma_x^i \sigma_x^{i+1}
\]
Here, I have treated i=1,N and i=2,3,...,N-1 separately.
The term \(-\frac{1}{2} \hcut g\sum_{i=2}^{N-1}   \sigma_x^i\) can be rewritten as,
\[
-\frac{1}{2} \hcut g\sum_{i=2}^{N-1}   \sigma_x^i = -\frac{1}{4} \hcut g\sum_{i=2}^{N-1}   \sigma_x^i-\frac{1}{4} \hcut g\sum_{i=1}^{N-2}   \sigma_x^{(i+1)}
\]
Here, both terms are equal but in the second term, I decrease the limits of summation by one, and increase the index of the summands by one.
Combining it with the other terms, we get,
\[
H_A =  - \frac{1}{4} \hcut g\sum_{i=1}^{N-1}   \sigma_x^i - \frac{1}{4} \hcut g\sum_{i=1}^{N-1}   \sigma_x^{i+1}  + \frac{1}{4} \hcut g\sum_{i=1}^{N-1}  \sigma_x^i \sigma_x^{i+1}
\]
Here, I added the i=1 term to the left summation and i=N to the right summation of the previous relation.

I can add another constant term,
\[\frac{1}{4} \hcut g\sum_{i=1}^{N-1}1\]
to the Hamiltonian, which will help us reach to our target form, without changing the dynamics of the system.
Then we get,
\[
H_A =  \frac{1}{4} \hcut g\sum_{i=1}^{N-1}1 - \frac{1}{4} \hcut g\sum_{i=1}^{N-1}   \sigma_x^i - \frac{1}{4} \hcut g\sum_{i=1}^{N-1}   \sigma_x^{i+1}  + \frac{1}{4} \hcut g\sum_{i=1}^{N-1}  \sigma_x^i\sigma_x^{i+1}
\]

Rearranging the terms we can write them as,

\[
H_A = \hbar g \sum_{i=1}^{N-1} \frac{1 - \sigma_x^i}{2} \frac{1 - \sigma_x^{i+1}}{2} \tag{8}
\]
Which is similar in form to that of the initial Hamiltonian.

\paragraph{Important Convention.}
We chose the states
\[
\ket{+}_i = \frac{\ket{0}_i - \ket{1}_i}{\sqrt{2}}, \quad \ket{-}_i = \frac{\ket{0}_i + \ket{1}_i}{\sqrt{2}}
\]
as the eigenstates of the Hamiltonian \( H_i = -\bar{E}_{J_i} \sigma^x_i \), with corresponding eigenvalues \( \pm\bar{E}_{J_i} \).

\medskip

\noindent
\textbf{This choice of eigenbasis is unconventional and specific to charge qubits.} From this point onwards, we will follow this convention throughout the discussion.

 \subsection{Projector Definition and Properties}

We introduce projectors as mathematical tools to analyze or simplify the Hamiltonian.
Now, define the projector:

\[
\Pi_i = \frac{1 - \sigma_i^x}{2}
\]

This projects onto the state \( |+\rangle_i = \frac{|0\rangle_i - |1\rangle_i}{\sqrt{2}} \), an eigenvector of \(\sigma_i^x\) with eigenvalue \(-1\):

\[
\sigma_i^x |+\rangle_i = -|+\rangle_i, \quad \sigma_i^x |-\rangle_i = |-\rangle_i, \quad |+\rangle_i = \frac{|0\rangle_i - |1\rangle_i}{\sqrt{2}}
\]

\[
\Pi_i |+\rangle_i = \frac{1 - (-1)}{2} |+\rangle_i = |-\rangle_i, \quad \Pi_i |-\rangle_i = \frac{1 - 1}{2} |-\rangle_i = 0
\]

Thus, \( \Pi_i = |+\rangle_i \langle +|_i \), and the two-qubit projector is:

\[
\Pi_i \Pi_{i+1} = |+\rangle_i |+\rangle_{i+1} \langle +|_i \langle +|_{i+1}
\]

Since \( (\Pi_i \Pi_{i+1})^2 = \Pi_i \Pi_{i+1} \), it is idempotent, a key property for computing its exponential.
Also, it has no imaginary part. So, it is hermitian, \(\Pi_i = \Pi_i^{\dagger} \).

\subsection{Unitary Evolution of the Hamiltonian}
Now, we can write the evolution operation corresponding to the Hamiltonian as follows,

\[
U(t) = \exp\left[ -i \frac{H_A t}{\hbar} \right]= \exp\left[ -i g t \sum_{i=1}^{N-1} \frac{1 - \sigma_i^x}{2} \frac{1 - \sigma_{i+1}^x}{2} \right]  \tag{9}
\]

Since \( H_A = \hbar g \sum_{i=1}^{N-1} \Pi_i \Pi_{i+1} \), and we need to check whether the terms \( \Pi_i \Pi_{i+1} \) commute, to factorize the Hamiltonian, and to consider the evolution locally(on each Qubit).
\subsubsection{Commutativity of Projectors}
Assume we are working on 3 qubits, labeled \(i = 0, 1, 2\). Define the following projectors (built from Pauli \(X\)):

\[
\Pi_0 = \frac{1}{2}(I - X_0) \otimes I_1 \otimes I_2
\]
\[
\Pi_1 = I_0 \otimes \frac{1}{2}(I - X_1) \otimes I_2
\]
\[
\Pi_2 = I_0 \otimes I_1 \otimes \frac{1}{2}(I - X_2)
\]

Now compute the product:

\[
\Pi_0 \Pi_1 = \left( \frac{1}{2}(I - X_0) \right) \otimes \left( \frac{1}{2}(I - X_1) \right) \otimes I_2
= \frac{1}{4}(I - X_0) \otimes (I - X_1) \otimes I
\]

Similarly:

\[
\Pi_1 \Pi_2 = I_0 \otimes \left( \frac{1}{2}(I - X_1) \right) \otimes \left( \frac{1}{2}(I - X_2) \right)
= \frac{1}{4} I \otimes (I - X_1) \otimes (I - X_2)
\]

We now compute the commutator:

\[
[\Pi_0 \Pi_1, \Pi_1 \Pi_2] = \Pi_0 \Pi_1 \Pi_1 \Pi_2 - \Pi_1 \Pi_2 \Pi_0 \Pi_1
\]

Note that \(\Pi_1\) is a projector, so \(\Pi_1^2 = \Pi_1\), and thus:

\[
= \Pi_0 \Pi_1 \Pi_2 - \Pi_1 \Pi_2 \Pi_0 \Pi_1
\]

We cannot simplify the expression further in general. This could lead us to use Trotterization\cite{Kluber_2023}(An approximate factorization of non-commuting Hamiltonians)

But fortunately, all the projectors here are simultaneously diagonalized by the x-basis, thus they commute with each other.

So, 
\[
[\Pi_0 \Pi_1, \Pi_1 \Pi_2] = 0
\]

% I have also, used the sympy, an open-source library for symbolic mathematics in Python, to cross-check the commutator. The code is given below,
% \begin{lstlisting}[style=python, caption={Commutator of adjacent projectors on 3 qubits}, label={lst:projector_commutator}]
% import sympy as sp
% from sympy.physics.quantum import TensorProduct as TP

% # Define 2x2 identity and Pauli-X matrix
% I = sp.eye(2)
% X = sp.Matrix([[0, 1], [1, 0]])

% # Define the single-qubit projector onto the -1 eigenspace of X
% def Pi(j):
%     return (I - X) / 2  # same for any j, used symbolically

% # Construct projectors on a 3-qubit system
% Pi0 = TP(Pi(0), I, I)
% Pi1 = TP(I, Pi(1), I)
% Pi2 = TP(I, I, Pi(2))

% # Compute the products of adjacent projectors
% Pi0Pi1 = Pi0 * Pi1
% Pi1Pi2 = Pi1 * Pi2

% # Compute the commutator [Pi0 Pi1, Pi1 Pi2]
% commutator = Pi0Pi1 * Pi1Pi2 - Pi1Pi2 * Pi0Pi1

% # Simplify and display result
% commutator = sp.simplify(commutator)
% sp.pprint(commutator)
% \end{lstlisting}

Now, we can factorize the Hamiltonian exactly.

% ------

\subsubsection{Factorizing the Hamiltonian}
The unitary factorizes:

\[
U(t) = \prod_{i=1}^{N-1} \ e^{\left[ -i g t \Pi_i \Pi_{i+1} \right]}
\]

For each term, \( \Pi_i \Pi_{i+1} = |+\rangle_i |+\rangle_{i+1} \langle +|_i \langle +|_{i+1} \) is a projector (\( (\Pi_i \Pi_{i+1})^2 = \Pi_i \Pi_{i+1} \)). For a projector \( A \), the exponential is:

\[
e^{-i \theta A} = I + (e^{-i \theta} - 1) A 
\] (see Appendix~\ref{appendix:projector_identity})

since \( A^k = A \) for \( k \geq 1 \). For \( \theta = g t \):

\[
e^{-i g t \Pi_i \Pi_{i+1}} = I + (e^{-i g t} - 1) \Pi_i \Pi_{i+1}
\]

Thus, for \( g t = (2n+1)\pi \):

\[
e^{-i (2n+1)\pi} = (-1)^{2n+1} = -1, \quad e^{-i \pi \Pi_i \Pi_{i+1}} = I + (-1 - 1) \Pi_i \Pi_{i+1} = I - 2 \Pi_i \Pi_{i+1}
\],
\[
 \quad U(t) = \prod_{i=1}^{N-1} \left[ I - 2 \Pi_i \Pi_{i+1} \right]
\]

Each factors apply a phase of \(-1\) to states where qubits \( i \) and \( i+1 \) are both \( |+\rangle \).

\subsubsection{Evolution of the System}
% Equation 10 states that for \( g t = (2n+1)\pi \), the evolved state is a cluster state:

% \[
% |N\rangle = \frac{1}{2^{N/2}} \prod_{i=1}^N \left[ |-\rangle_i + |+\rangle_i \sigma_{i+1}^x \right], \quad \sigma_{N+1}^x \equiv 1 \tag{10}
% \]

The initial state is:

\[
|\psi_0\rangle = \bigotimes_{i=1}^N |0\rangle_i
\]

where \( |0\rangle_i \) is prepared with \( C_i V_i / e \approx 0 \). At this configuration and as we assume that the charge qubit works in the charging regime with \(E_{ci} \gg
E_{Ji}\), the effective single qubit hamiltonian is \(H_i = \varepsilon_i(V_i) \sigma_z^{(i)}\). The evolution of this Hamiltonian simply adds global phases to the initial state\( |0\rangle_i \), with zero excess Cooper pair.  Express \( |0\rangle_i \) in the \( \sigma_i^x \) basis:

\[
|0\rangle_i = \frac{|-\rangle_i + |+\rangle_i}{\sqrt{2}}\] 
where, \( |+\rangle_i = \frac{|0\rangle_i - |1\rangle_i}{\sqrt{2}}, \quad |-\rangle_i = \frac{|0\rangle_i + |1\rangle_i}{\sqrt{2}}
\)

since \( \sigma_i^x |+\rangle_i = -|+\rangle_i \), \( \sigma_i^x |-\rangle_i = |-\rangle_i \). Thus:

\[
|\psi_0\rangle = \bigotimes_{i=1}^N \frac{|-\rangle_i + |+\rangle_i}{\sqrt{2}} \]

% = \frac{1}{2^{N/2}} \prod_{i=1}^N \left[ |-\rangle_i + |+\rangle_i \right]
% \]

\noindent The evolved state is:

\[
|\psi\rangle = U(t) |\psi_0\rangle = \prod_{i=1}^{N-1} \left[ I - 2 \Pi_i \Pi_{i+1} \right] |\psi_0\rangle
\]

Now, let's denote the two‐qubit evolution block
\[
U_{i,i+1}
\;=\; I - 2\,\Pi_i\Pi_{i+1}
\;=\; I \;-\; 2\,\ket{+}_i\ket{+}_{i+1}\bra{+}_i\bra{+}_{i+1}.
\]

\medskip

\noindent Action on the \(X\)-basis:
\[
\begin{aligned}
U_{i,i+1}\ket{-}_i\ket{-}_{i+1} &= \ket{-}_i\ket{-}_{i+1},\\
U_{i,i+1}\ket{-}_i\ket{+}_{i+1} &= \ket{-}_i\ket{+}_{i+1},\\
U_{i,i+1}\ket{+}_i\ket{-}_{i+1} &= \ket{+}_i\ket{-}_{i+1},\\
U_{i,i+1}\ket{+}_i\ket{+}_{i+1} &= -\,\ket{+}_i\ket{+}_{i+1}.
\end{aligned}
\]

\medskip

\noindent Before and after using \(U_{i,i+1}\):,
\[
\begin{aligned}
&(\ket{-}_i+\ket{+}_i)(\ket{-}_{i+1}+\ket{+}_{i+1})
= \ket{-}_i\ket{-}_{i+1} + \ket{-}_i\ket{+}_{i+1} + \ket{+}_i\ket{-}_{i+1} + \ket{+}_i\ket{+}_{i+1},\\
&U_{i,i+1}\,\bigl[\ket{-}_i\ket{-}_{i+1} + \ket{-}_i\ket{+}_{i+1} + \ket{+}_i\ket{-}_{i+1} + \ket{+}_i\ket{+}_{i+1}\bigr] \\
&\quad= \ket{-}_i\ket{-}_{i+1} + \ket{-}_i\ket{+}_{i+1} + \ket{+}_i\ket{-}_{i+1} - \ket{+}_i\ket{+}_{i+1}.
\end{aligned}
\]

\medskip

\noindent Rewriting the results,
\[
\ket{-}_i\ket{-}_{i+1} + \ket{-}_i\ket{+}_{i+1} + \ket{+}_i\ket{-}_{i+1} - \ket{+}_i\ket{+}_{i+1}
= \bigl(\ket{-}_i + \ket{+}_i\,\sigma^x_{i+1}\bigr)
  \bigl(\ket{-}_{i+1} + \ket{+}_{i+1}\bigr),
\]
since \(\sigma^x_{i+1}\ket{+}_{i+1} = -\ket{+}_{i+1}\) and \(\sigma^x_{i+1}\ket{-}_{i+1} = \ket{-}_{i+1}\).
% We adopt the \(\sigmaX\) operator in our equation because, at the degeneracy point(the working condition of our system evolution), the \(\sigmaX\) operator serves as the sole generator of the system dynamics.
\medskip

\noindent If we consider all \(N\)-qubits, it will give us similar results(just change i into i+1 and i+1 into i+2, then we will get to the next pair of qubits), and resultant state will become
\[
\ket{\psi}
= \prod_{i=1}^{N-1} U_{i,i+1}\,\ket{\psi_0}
= \frac1{2^{N/2}}
\bigotimes_{i=1}^N\Bigl(\ket{-}_i + \ket{+}_i\,\sigma^x_{i+1}\Bigr),
\quad \sigma^x_{N+1} \equiv I, \tag{10}
\]
which is precisely our desired cluster state! We can rename \(\ket{\psi}\) to \(\ket{N}\).

% ---

\subsection{Verifying The Cluster State}
We first check, how it is related to the standard definition of cluster state. Then, we can immediately use the stabilizer proof to verify the cluster state. We ignore it, as it is directly evident after we show equivalence of the two forms. 
\subsection*{Verification}
We start from our final $X$‐basis form of the $N$‐qubit state:
\[
\ket{\Psi}
=\frac1{2^{N/2}}
\prod_{i=1}^N\bigl(\ket{-}_i+\ket{+}_i\,\sigma^x_{i+1}\bigr),
\quad\sigma^x_{N+1}\equiv I.
\]

\bigskip

\noindent Let's focus on one pair (i,i+1),denoted as \(F_{i,i+1}\):
\[
F_{i,i+1}
=(\ket{-}_i+\ket{+}_i\,\sigma^x_{i+1})
(\ket{-}_{i+1}+\ket{+}_{i+1}).
\]

\noindent Before I address the problem, let's have a quick refresher on the basics.
\bigskip

\noindent The single‐qubit Hadamard operator in matrix form is
\[
H=\frac1{\sqrt2}
\begin{pmatrix}
1 & 1\\
1 & -1
\end{pmatrix},
\]
\[
\ket{+} = \frac{\ket{0} - \ket{1}}{\sqrt{2}}=
\frac{1}{\sqrt{2}}\begin{pmatrix}1 \\ -1\end{pmatrix},
\quad \ket{-} = \frac{\ket{0} + \ket{1}}{\sqrt{2}}=\frac{1}{\sqrt{2}}\begin{pmatrix}1 \\ 1\end{pmatrix}
\quad\]

Writing our ket vectors in matrix form.

\noindent Hence, explicitly,
\[
H\ket{+}
= \frac{1}{\sqrt2}
\begin{pmatrix}
1 & 1 \\[3pt]
1 & -1
\end{pmatrix}\frac{1}{\sqrt{2}}
\begin{pmatrix}1 \\[3pt] -1\end{pmatrix}
= \begin{pmatrix}0 \\[3pt] 1\end{pmatrix}
\;=\;
\ket{1},
\]
\[
H\ket{-}
= \frac{1}{\sqrt2}
\begin{pmatrix}
1 & 1 \\[3pt]
1 & -1
\end{pmatrix}
\frac{1}{\sqrt{2}}
\begin{pmatrix}1 \\[3pt] 1\end{pmatrix}
= 
\begin{pmatrix}1 \\[3pt] 0\end{pmatrix}
\;=\;
\ket{0}.
\]

\medskip

\noindent\textbf{Conjugation of operators:} For any operator \(A\) and matrix \(M\),
\[
M \;:\; A \;\longmapsto\; M\,A\,M^\dagger.
\]
Since \(H^\dagger = H\), we compute
\[
H\,\sigma^x\,H
= \frac{1}{2}
\begin{pmatrix}
1 & 1 \\[3pt]
1 & -1
\end{pmatrix}
\begin{pmatrix}
0 & 1 \\[3pt]
1 & 0
\end{pmatrix}
\begin{pmatrix}
1 & 1 \\[3pt]
1 & -1
\end{pmatrix}
= \begin{pmatrix}
1 & 0 \\[3pt]
0 & -1
\end{pmatrix}
= \sigma^z,
\]
\[
H\,\sigma^z\,H
= \frac{1}{2}
\begin{pmatrix}
1 & 1 \\[3pt]
1 & -1
\end{pmatrix}
\begin{pmatrix}
1 & 0 \\[3pt]
0 & -1
\end{pmatrix}
\begin{pmatrix}
1 & 1 \\[3pt]
1 & -1
\end{pmatrix}
= \begin{pmatrix}
0 & 1 \\[3pt]
1 & 0
\end{pmatrix}
= \sigma^x.
\]

\bigskip

\noindent\textbf{Basis Change in Quantum Measurements}

When performing quantum measurements in different bases, we must transform both the quantum state and the applied operators consistently. For a unitary transformation $U$, the fundamental rule requires that we simultaneously apply $U$ to the state and conjugate all operators by $U$. We can achieve the same result by first allowing all quantum operations to act on the initial state (representing this cumulative effect as operator A), then applying the unitary transformation that rotates from the standard computational basis to our desired measurement basis. Starting with the later definition,

$$|\psi\rangle=A|\psi_0\rangle \xrightarrow{U} UA|\psi_0\rangle=UAI|\psi_0\rangle=(UAU^\dagger)(U|\psi_0\rangle)$$

This establishes the equivalence of both description. 
\medskip

\noindent\textbf{Transformation from Z Basis to X Basis}

The transformation from the computational Z basis to the X basis utilizes the Hadamard gate $H$. A vector in the $z$-basis is transformed into a $x$-basis(\(|+\rangle,|-\rangle\) vector and vice versa under the action of $H$ gate.

When we want to measure in $x$-basis on a state $|\psi\rangle$, which is transformed by $\sigmaX$, we can equivalently write:

\[\sigma^x|\psi\rangle \xrightarrow{H} H\sigma^x|\psi\rangle = H \sigma^x H \cdot H|\psi\rangle = \sigma^z H|\psi\rangle \tag{i}\]

\bigskip

\noindent Now let's try changing the measurement basis from $z$-basis to $x$-basis to see whether we find any pattern that matches the standard cluster state form. 

\bigskip
\noindent We have seen in equation (i) , measuring in $x$ basis implies that, we use a Hadamard to all qubits and conjugate the operators.

We transform each qubit: $\ket{+}\xrightarrow{H}\ket0$, $\ket{-}\xrightarrow{H}\ket1$, and
$\sigma^x_{i+1}\xrightarrow{H (\cdot) H^{-1}} H \sigma^x_{i+1} H^{-1}=
\sigma^z_{i+1}$.  Thus
\[
H_iH_{i+1}\,F_{i,i+1}
=\bigl(\ket{1}_i+\ket{0}_i\,\sigma^z_{i+1}\bigr)
\bigl(\ket{1}_{i+1}+\ket{0}_{i+1}\bigr).
\]
Reorder terms:
\[
=\bigl(\ket{0}_i+\ket{1}_i\,\sigma^z_{i+1}\bigr)
\bigl(\ket{0}_{i+1}+\ket{1}_{i+1}\bigr).
\]

\bigskip
\noindent If we look closely, we will recognize CZ action.

Expanding in the computational basis:
\[
\begin{aligned}
&(\ket{0}_i + \ket{1}_i\,\sigma^z_{i+1})(\ket{0}_{i+1} + \ket{1}_{i+1}) \\
&\quad= \ket{0}_i\ket{0}_{i+1} + \ket{0}_i\ket{1}_{i+1}
       + \ket{1}_i\,\sigma^z_{i+1}\ket{0}_{i+1}
       + \ket{1}_i\,\sigma^z_{i+1}\ket{1}_{i+1} \\
&\quad= \ket{0}_i\ket{0}_{i+1} + \ket{0}_i\ket{1}_{i+1}
       + \ket{1}_i\ket{0}_{i+1} - \ket{1}_i\ket{1}_{i+1} \\
&\quad= \text{CZ}_{i,i+1} \big(\ket{0}_i\ket{0}_{i+1}
       + \ket{0}_i\ket{1}_{i+1}
       + \ket{1}_i\ket{0}_{i+1}
       + \ket{1}_i\ket{1}_{i+1} \big) \\
&\quad= \text{CZ}_{i,i+1} \big(\ket{+}_i\ket{+}_{i+1}\big).
\end{aligned}
\]

Here $\text{CZ}=\mathrm{diag}(1,1,1,-1)$ introduces the minus on $\ket{11}$.

\bigskip
\noindent Since each pair $(i,i+1)$ satisfies
\[
H^{\otimes 2} F_{i,i+1}
= \text{CZ}_{i,i+1} \ket{+}_i\ket{+}_{i+1}
\]

and these CZ gates commute on different pairs, applying $H^{\otimes N}$
to the full product yields
\[
H^{\otimes N}\,\ket{\Psi}
=\prod_{i=1}^{N-1}\text{CZ}_{i,i+1}\;\ket{+}^{\otimes N}
\;=\;\ket{C_N},
\]

\medskip

\medskip

This is the standard 1D or linear cluster state.

\section{Simulation Methodology}
\label{sec:simulation}

This section outlines the methodology employed to simulate the Hamiltonian dynamics of an \( n \)-qubit system aimed at generating a cluster state, a universal resource for measurement-based quantum computing (MBQC)~\cite{Raussendorf2001}. The simulation was conducted using QuTiP, a Python framework for modeling open quantum systems~\cite{qutip2024}. We detail the simulation framework, fidelity calculation, and the incorporation of decoherence effects characterized by T1 and T2 coherence times, with results reserved for the subsequent section.

\subsection{QuTiP: A Framework for Quantum Simulations}
QuTiP (Quantum Toolbox in Python) is an open-source framework designed for simulating the dynamics of quantum systems, particularly those subject to environmental interactions~\cite{qutip2024}. It provides robust tools for solving the Lindblad master equation, which governs the evolution of a quantum system under both unitary dynamics (via the Hamiltonian) and dissipative processes (such as decoherence). In this work, QuTiP’s Lindblad master-equation solver was utilized to model the time evolution of an \( n \)-qubit system under a Hamiltonian engineered to produce a cluster state. The solver’s ability to handle both ideal and noisy dynamics makes it well-suited for evaluating the feasibility of cluster state generation in realistic quantum hardware.

\subsection{Fidelity: Definition and Computation}
Fidelity is a fundamental metric in quantum information science, quantifying the similarity between two quantum states~\cite{Nielsen_Chuang_2010}. It is used here to assess how closely the simulated state \( \rho_{\text{sim}}(t) \) at time \( t \) matches the ideal cluster state \( \sigma_{\text{cluster}} = \ket{\psi_{\text{cluster}}}\bra{\psi_{\text{cluster}}} \). For pure states, fidelity is defined as:

\begin{equation}
F(t) = \left| \langle \psi_{\text{cluster}} | \psi_{\text{sim}}(t) \rangle \right|^2.
\end{equation}

In the presence of decoherence, the simulated state may become mixed, requiring the generalized fidelity for mixed states\cite{Jozsa1994Fidelity}:

\begin{equation}
F(t) = \left( \mathrm{Tr} \sqrt{ \sqrt{\rho_{\text{sim}}(t)} \sigma_{\text{cluster}} \sqrt{\rho_{\text{sim}}(t)} } \right)^2.
\end{equation}

This formulation ensures accurate fidelity computation under noisy conditions, leveraging QuTiP’s built-in \texttt{fidelity} function, which efficiently handles both pure and mixed states~\cite{qutip2024}. Fidelity serves as a critical measure for verifying the success of cluster state generation, with higher values indicating greater overlap with the target state.

\subsection{Decoherence Effects: T1 and T2}
Quantum systems are inherently susceptible to decoherence, which degrades quantum information over time. The two primary decoherence mechanisms are:

\begin{itemize}
    \item \textbf{T1 Relaxation}: The process by which a qubit in the excited state \( \ket{1} \) decays to the ground state \( \ket{0} \), characterized by the relaxation time T1. This energy loss reduces the qubit’s ability to maintain superposition.
    \item \textbf{T2 Dephasing}: The loss of phase coherence between \( \ket{0} \) and \( \ket{1} \), leading to a mixed state, characterized by the dephasing time T2. This is caused by environmental fluctuations disrupting phase relationships.
\end{itemize}

In our simulation, T1 and T2 effects were modeled using QuTiP’s Lindblad operators, which describe dissipative processes in the master equation. 
% Specifically:

% \begin{itemize}
%     \item \textbf{Relaxation (T1)}: Represented by the Lindblad operator \( \mathcal{L}_{\text{relax}} = \sqrt{\gamma_1} \sigma^- \), where \( \gamma_1 = 1 / T1 \) and \( \sigma^- = \ket{0}\bra{1} \) is the lowering operator.
%     \item \textbf{Dephasing (T2)}: Represented by the Lindblad operator \( \mathcal{L}_{\text{dephase}} = \sqrt{\gamma_2 / 2} \sigma^z \), where \( \gamma_2 = 1 / T2 - \gamma_1 / 2 \) and \( \sigma^z \) is the Pauli Z operator.
% \end{itemize}

% These operators were incorporated into the Lindblad master equation to simulate the impact of decoherence on cluster state generation. 
We adopted median coherence times from IBM’s latest transmon charge qubits, with \( T_1 = 262.69\)\(\mu s\) and \( T_2 = 176.67\)\(\mu s\)~\cite{AbuGhanem2024}. These values reflect state-of-the-art superconducting qubit performance. 

\section{Results}

Instead of general n, we selected n = 4‑qubit cluster state. We compare our expectations—unity peaks at odd multiples of $\pi$ under ideal evolution—with the simulation outcomes when including decoherence.

\subsection{Ideal Evolution}
Under the noise‑free Hamiltonian, the fidelity reaches near 100\% at each anticipated revival (odd multiples of $\pi$), confirming correct implementation of the cluster‐state generator.
\begin{figure}[ht]
\centering
\includegraphics[width=0.7\linewidth]{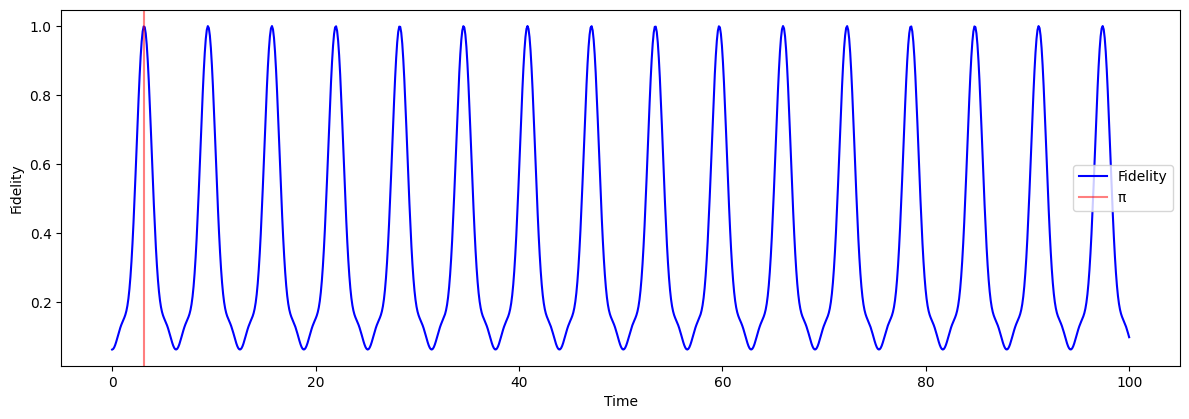}
\caption{Ideal evolution fidelity of The 4-qubit Cluster State Generating Hamiltonian.}
\label{fig:ideal_fidelity}
\end{figure}

\subsection{T1 Relaxation}
With only \(T_1\) processes active, we observe that each fidelity peak is reduced compared to the ideal case. The peak at \(t=\pi\) remains above 90\%, but by the fourth revival it falls to approximately 80\%, illustrating population decay in the excited manifold.
\begin{figure}[ht]
\centering
\includegraphics[width=0.7\linewidth]{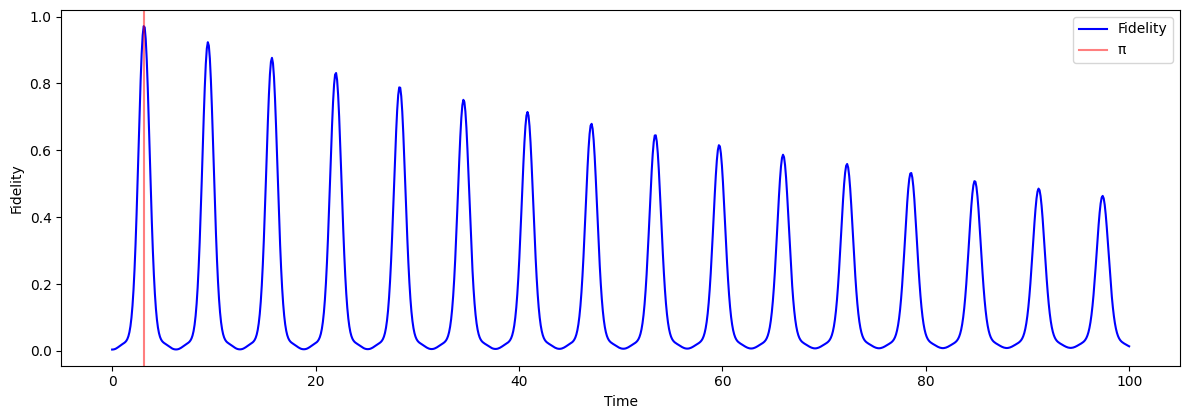}
\caption{\(T_1\) decoherence fidelity for 4-Qubit Cluster State Generation Hamiltonian}
\label{fig:T1_fidelity}
\end{figure}

\subsection{T2 Dephasing}
Introducing pure dephasing \(T_2\) leads to a faster drop in oscillation contrast. While the first peak still exceeds 90\%, subsequent revivals decline to around 70\% by the fourth. This matches our expectation that loss of phase coherence is more detrimental than energy decay alone for this protocol.
\begin{figure}[ht]
\centering
\includegraphics[width=0.7\linewidth]{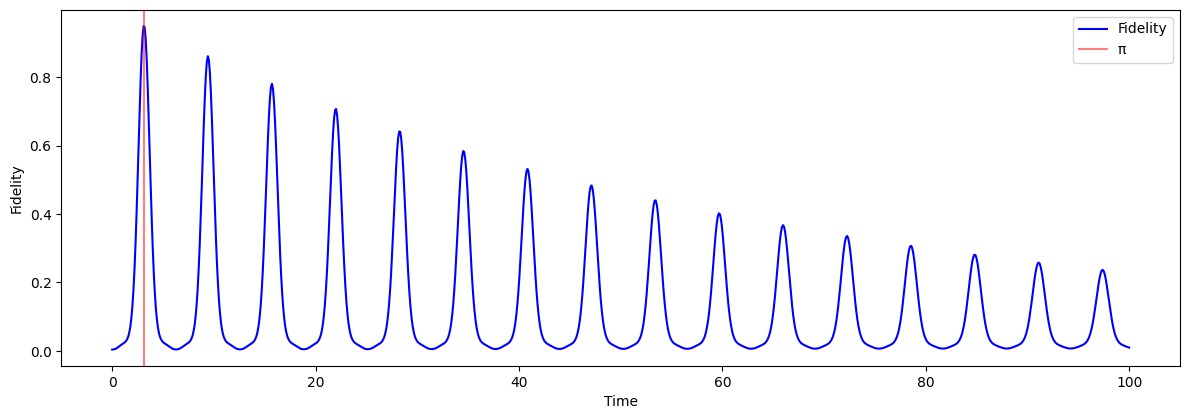}
\caption{\(T_2\) decoherence fidelity plot of 4-qubit Cluster State Generation}
\label{fig:T2_fidelity}
\end{figure}

\subsection{Combined Decoherence}
When both \(T_1\) and \(T_2\) act together, fidelity peaks shrink most dramatically: the first revival is near 85\%, and later peaks fall below 70\%. This worst‐case scenario underscores the challenge of cluster‐state preparation on real hardware.
\begin{figure}[ht]
\centering
\includegraphics[width=0.7\linewidth]{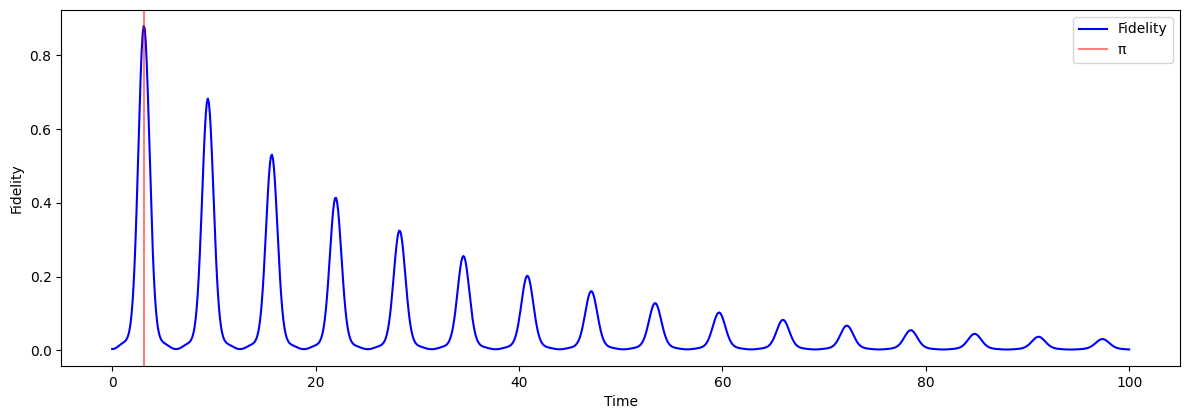}
\caption{Combined fidelity plot of 4-qubit Cluster State Generation}
\label{fig:combined_fidelity}
\end{figure}

\subsection{Coherence After Projection}
By sampling the state at \(t=\pi\) and then tracking its total off‑diagonal coherence, we find a roughly exponential decay: under combined decoherence, coherence drops to 50\% within 15 time units after projection, whereas with only \(T_1\) it remains above 70\% over the same interval.

\begin{figure}[H]
\centering
\includegraphics[width=0.7\linewidth]{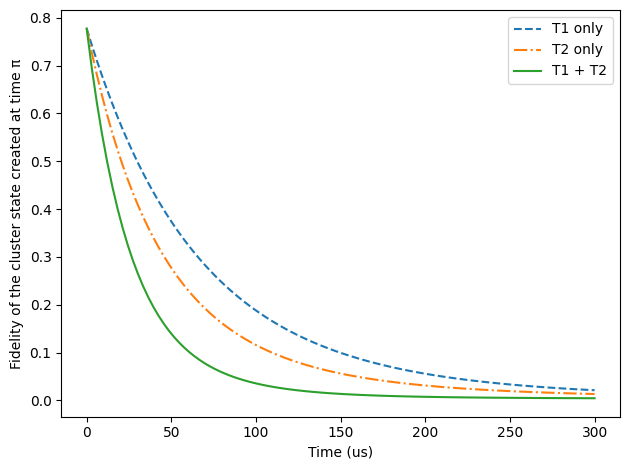}
\caption{Coherence‐decay plot for 4-qubit Cluster State Generated at time $\pi$}
\label{fig:first_cs_fidelity}
\end{figure}

Overall, the results align with theoretical predictions: ideal revivals are perfect, \(T_2\) degrades peaks faster than \(T_1\), and the combination of channels poses the greatest limitation. Quantitatively, the 4‑qubit system remains demonstrably coherent over a few revival periods, but scaling to larger clusters will require mitigation of both relaxation and dephasing.

\section{Discussions and Future Work}
Our simulations demonstrate that cluster states can be generated through Hamiltonian evolution with high fidelity when operations are performed at optimal times \(t = (2n+1)\pi\). However, decoherence, particularly \(T_2\) dephasing, presents significant limitations by degrading entanglement after the first fidelity peak at \(t = \pi\). This highlights the critical need to synchronize measurement-based quantum computing (MBQC) protocols with narrow windows of high coherence. Such timing constraints are consistent with observations in current superconducting qubit platforms.
Several directions for future research emerge from this work. Alternative Hamiltonian descriptions could be investigated, particularly the Jaynes-Cummings interaction\cite{bina2012coherent}, to determine whether different coupling regimes produce broader or more robust fidelity peaks. Implementations beyond superconducting circuits warrant exploration, including color-center systems in diamond (NV centers)\cite{casanova2016noise} and silicon-vacancy qubits in SiC\cite{castelletto2020silicon}. These studies would assess how intrinsic material properties and noise characteristics affect cluster-state generation. Error-mitigation strategies could be integrated to sustain coherence for practical MBQC applications. Such strategies include dynamical decoupling sequences and small-scale quantum error-correcting codes. These research directions aim to bridge the gap between theoretical cluster-state frameworks and the practical constraints of noisy quantum hardware systems.

\section{Conclusions}
\label{sec:conclusions}

This study explores the generation of an \( n \)-qubit cluster state for measurement-based quantum computing (MBQC). We verified the cluster state, achieving our main goal. The simulations showed that the cluster state forms with high fidelity at times \( t = (2n+1)\pi \) without noise, confirming the Hamiltonian works as intended.

When we added T\(_1\) (\SI{262.69}{\micro\second}) and T\(_2\) (\SI{176.67}{\micro\second}) decoherence, based on IBM’s transmon qubits~\cite{AbuGhanem2024}, the fidelity peaks dropped over time. This shows noise, especially T\(_2\), harms the cluster state’s quality. At \( t = \pi \), the state still had good fidelity despite noise, but its coherence fell quickly after, meaning MBQC needs fast operations to work well.

These results shows the noise response of cluster state generated using superconducting charge qubits. Future work could try to engineer the hardware to make noise resilient, robust hardware considering the goal to generate cluster states.

\printbibliography

@article{PhysRevA.75.052319,
  title   = {Efficient one-step generation of large cluster states with solid-state circuits},
  author  = {You, J. Q. and Wang, Xiang-bin and Tanamoto, Tetsufumi and Nori, Franco},
  journal = {Phys. Rev. A},
  volume  = {75},
  number  = {5},
  pages   = {052319},
  year    = {2007},
  doi     = {10.1103/PhysRevA.75.052319}
}

@book{Nielsen_Chuang_2010,
  author    = {Nielsen, Michael A. and Chuang, Isaac L.},
  title     = {Quantum Computation and Quantum Information: 10th Anniversary Edition},
  publisher = {Cambridge University Press},
  address   = {Cambridge},
  year      = {2010},
  doi       = {10.1017/CBO9780511976667}
}

@article{Kluber_2023,
  title={Trotterization in Quantum Theory}, 
      author={Physics Claire Kluber},
      year={2025},
      eprint={2310.13296},
      archivePrefix={arXiv},
      primaryClass={quant-ph},
      url={https://arxiv.org/abs/2310.13296}, 
}

@article{lanyon2008measurement,
  title = {Measurement-Based Quantum Computation with Trapped Ions},
  author = {Lanyon, B. P. and Jurcevic, P. and Zwerger, M. and Hempel, C. and Martinez, E. A. and D\"ur, W. and Briegel, H. J. and Blatt, R. and Roos, C. F.},
  journal = {Phys. Rev. Lett.},
  volume = {111},
  issue = {21},
  pages = {210501},
  numpages = {5},
  year = {2013},
  month = {11},
  publisher = {American Physical Society},
  doi = {10.1103/PhysRevLett.111.210501},
  url = {https://link.aps.org/doi/10.1103/PhysRevLett.111.210501}
}

@article{Raussendorf2001,
  author  = {Raussendorf, Robert and Briegel, Hans J.},
  title   = {A One-Way Quantum Computer},
  journal = {Phys. Rev. Lett.},
  volume  = {86},
  number  = {22},
  pages   = {5188--5191},
  year    = {2001},
  doi     = {10.1103/PhysRevLett.86.5188}
}

@article{Jozsa1994Fidelity,
  author  = {Jozsa, Richard},
  title   = {Fidelity for Mixed Quantum States},
  journal = {J. Mod. Opt.},
  volume  = {41},
  number  = {12},
  pages   = {2315--2323},
  year    = {1994},
  doi     = {10.1080/09500349414552171}
}

@misc{qutip2024,
  title         = {QuTiP: Quantum Toolbox in Python},
  author        = {Alexander, R. N. and Burkard, Guido and Criger, Ben and Dassonneville, Romain and del Rey, Maike and DiVincenzo, David P. and others},
  eprint        = {2412.04705},
  archivePrefix = {arXiv},
  primaryClass  = {quant-ph},
  year          = {2024}
}

@article{shor1997,
  author  = {Shor, Peter W.},
  title   = {Polynomial-Time Algorithms for Prime Factorization and Discrete Logarithms on a Quantum Computer},
  journal = {SIAM J. Comput.},
  volume  = {26},
  number  = {5},
  pages   = {1484--1509},
  year    = {1997},
  doi     = {10.1137/S0097539795293172}
}

@article{feynman1982,
  author  = {Feynman, Richard P.},
  title   = {Simulating physics with computers},
  journal = {Int. J. Theor. Phys.},
  volume  = {21},
  number  = {6--7},
  pages   = {467--488},
  year    = {1982},
  doi     = {10.1007/BF02650179}
}

@article{devoret2013superconducting,
  author  = {Devoret, Michel H. and Schoelkopf, Robert J.},
  title   = {Superconducting Circuits for Quantum Information: An Outlook},
  journal = {Science},
  volume  = {339},
  number  = {6124},
  pages   = {1169--1174},
  year    = {2013},
  doi     = {10.1126/science.1231930}
}

@article{briegel2009measurement,
  author  = {Briegel, Hans J. and Browne, Dan E. and D{\"u}r, Wolfgang and Raussendorf, Robert and Van den Nest, Maarten},
  title   = {Measurement-based quantum computation},
  journal = {Nat. Phys.},
  volume  = {5},
  number  = {1},
  pages   = {19--26},
  year    = {2009},
  doi     = {10.1038/nphys1157}
}

@article{tame2008measurement,
  title = {Measurement-Based Quantum Computation with Trapped Ions},
  author = {Lanyon, B. P. and Jurcevic, P. and Zwerger, M. and Hempel, C. and Martinez, E. A. and D\"ur, W. and Briegel, H. J. and Blatt, R. and Roos, C. F.},
  journal = {Phys. Rev. Lett.},
  volume = {111},
  issue = {21},
  pages = {210501},
  numpages = {5},
  year = {2013},
  month = {11},
  publisher = {American Physical Society},
  doi = {10.1103/PhysRevLett.111.210501},
  url = {https://link.aps.org/doi/10.1103/PhysRevLett.111.210501}
}

@article{preskill2018quantum,
  author  = {Preskill, John},
  title   = {Quantum Computing in the NISQ Era and Beyond},
  journal = {Quantum},
  volume  = {2},
  pages   = {79},
  year    = {2018},
  doi     = {10.22331/q-2018-08-06-79}
}

@article{Clarke2008,
  author  = {Clarke, John and Wilhelm, Frank K.},
  title   = {Superconducting quantum bits},
  journal = {Nature},
  volume  = {453},
  number  = {7198},
  pages   = {1031--1042},
  year    = {2008},
  doi     = {10.1038/nature07128}
}

@phdthesis{Gottesman1997,
  author = {Gottesman, Daniel},
  title  = {Stabilizer Codes and Quantum Error Correction},
  school = {California Institute of Technology},
  year   = {1997},
  type   = {Ph.D. thesis},
  url={https://arxiv.org/abs/quant-ph/9705052}
}

@article{VanDenNest2007,
  author  = {Van den Nest, Maarten and Dehaene, Jeroen and De Moor, Bart},
  title   = {Efficient algorithm to recognize the local equivalence of graph states and compute their classical correlations},
  journal = {Phys. Rev. A},
  volume  = {75},
  number  = {1},
  pages   = {012312},
  year    = {2007},
  doi     = {10.1103/PhysRevA.75.012312}
}

@article{Koch2007,
  author  = {Koch, Jens and Yu, Terri M. and Gambetta, Jay and Houck, A. A. and Schuster, D. I. and Majer, J. and Blais, Alexandre and Devoret, M. H. and Girvin, S. M. and Schoelkopf, R. J.},
  title   = {Charge-insensitive qubit design derived from the Cooper pair box},
  journal = {Phys. Rev. A},
  volume  = {76},
  number  = {4},
  pages   = {042319},
  year    = {2007},
  doi     = {10.1103/PhysRevA.76.042319}
}

@article{Makhlin2001,
  author  = {Makhlin, Yuriy and Sch{\"o}n, Gerd and Shnirman, Alexander},
  title   = {Quantum-state engineering with Josephson-junction devices},
  journal = {Rev. Mod. Phys.},
  volume  = {73},
  number  = {2},
  pages   = {357--400},
  year    = {2001},
  doi     = {10.1103/RevModPhys.73.357}
}

@article{Martinis2002,
  author  = {Martinis, John M. and Nam, S. and Aumentado, J. and Urbina, C.},
  title   = {Rabi oscillations in a large Josephson-junction qubit},
  journal = {Phys. Rev. Lett.},
  volume  = {89},
  number  = {11},
  pages   = {117901},
  year    = {2002},
  doi     = {10.1103/PhysRevLett.89.117901}
}

@article{AbuGhanem2024,
  author  = {AbuGhanem, M.},
  title   = {IBM Quantum Computers: Evolution, Performance, and Future Directions},
  journal = {arXiv preprint arXiv:2410.00916},
  year    = {2024},
  doi     = {10.48550/arXiv.2410.00916}
}

@article{castelletto2020silicon,
  author    = {Castelletto, Stefania and Boretti, Alberto},
  title     = {Silicon carbide color centers for quantum applications},
  journal   = {J. Phys.: Photonics},
  volume    = {2},
  number    = {2},
  pages     = {022001},
  year      = {2020},
  doi       = {10.1088/2515-7647/ab77a2}
}

@article{casanova2016noise,
  title = {Noise-Resilient Quantum Computing with a Nitrogen-Vacancy Center and Nuclear Spins},
  author = {Casanova, J. and Wang, Z.-Y. and Plenio, M. B.},
  journal = {Phys. Rev. Lett.},
  volume = {117},
  issue = {13},
  pages = {130502},
  numpages = {6},
  year = {2016},
  month = {09},
  publisher = {American Physical Society},
  doi = {10.1103/PhysRevLett.117.130502},
  url = {https://link.aps.org/doi/10.1103/PhysRevLett.117.130502}
}

@article{bina2012coherent,
  author        = {Bina, Matteo},
  title         = {The coherent interaction between matter and radiation: A tutorial on the Jaynes--Cummings model},
  journal       = {Eur. Phys. J. Spec. Top.},
  volume        = {203},
  pages         = {163--183},
  year          = {2012},
  doi           = {10.1140/epjst/e2012-01541-3},
  eprint        = {1111.1143},
  archivePrefix = {arXiv},
  primaryClass  = {quant-ph}
}

@article{RevModPhys.90.015002,
  author  = {Albash, Tameem and Lidar, Daniel A.},
  title   = {Adiabatic quantum computation},
  journal = {Rev. Mod. Phys.},
  volume  = {90},
  number  = {1},
  pages   = {015002},
  year    = {2018},
  doi     = {10.1103/RevModPhys.90.015002}
}

@misc{freedman2002topologicalquantumcomputation,
  author        = {Freedman, Michael H. and Kitaev, Alexei and Larsen, Michael J. and Wang, Zhenghan},
  title         = {Topological Quantum Computation},
  eprint        = {quant-ph/0101025},
  archivePrefix = {arXiv},
  primaryClass  = {quant-ph},
  year          = {2002}
}

@article{Venegas_Andraca_2012,
  author  = {Venegas-Andraca, Salvador Elías},
  title   = {Quantum walks: a comprehensive review},
  journal = {Quantum Inf. Process.},
  volume  = {11},
  number  = {5},
  pages   = {1015--1106},
  year    = {2012},
  doi     = {10.1007/s11128-012-0432-5}
}

@article{Schreier2007SuppressingCN,
  author  = {Schreier, J. A. and Houck, Andrew A. and Koch, Jens and Schuster, David I. and Johnson, Blake R. and Chow, Jerry M. and Gambetta, Jay M. and Majer, Johannes and Frunzio, Luigi and Devoret, Michel H. and Girvin, Steven M. and Schoelkopf, Robert J.},
  title   = {Suppressing charge noise decoherence in superconducting charge qubits},
  journal = {Phys. Rev. B},
  volume  = {77},
  pages   = {180502},
  year    = {2008},
  doi     = {10.1103/PhysRevB.77.180502}
}

@inproceedings{Li2025,
author = {Li, Yingheng and Pawar, Aditya and Azari, Mohadeseh and Guo, Yanan and Zhang, Youtao and Yang, Jun and Seshadreesan, Kaushik Parasuram and Tang, Xulong},
title = {Orchestrating Measurement-Based Quantum Computation over Photonic Quantum Processors},
year = {2025},
isbn = {9798350323481},
publisher = {IEEE Press},
url = {https://doi.org/10.1109/DAC56929.2023.10247944},
doi = {10.1109/DAC56929.2023.10247944},
booktitle = {Proceedings of the 60th Annual ACM/IEEE Design Automation Conference},
pages = {1–6},
numpages = {6},
location = {San Francisco, California, United States},
series = {DAC '23}
}

@misc{Graham2024,
  author = {Thomas M. Graham and Mark Saffman},
  title = {Fast Feedback for Measurement-Based Quantum Computation with Neutral Atoms},
  year = {2024},
  url = {https://purl.stanford.edu/py978gh8681}
}

\appendix
\section*{Appendix}

% Derivation of \texorpdfstring{$e^{-i \theta A} = I + (e^{-i \theta} - 1) A$}{exp(-iθA) identity}}
\subsection*{1. Derivation of \texorpdfstring{$e^{-i \theta A} = I + (e^{-i \theta} - 1) A$}{exp(-iθA) identity}}
\label{appendix:projector_identity}
Let \( A \) be a projector, i.e., \( A^2 = A \). We aim to evaluate the exponential \( e^{-i\theta A} \).
Recall the power series expansion of the exponential function:
\[
e^{-i \theta A} = \sum_{n=0}^{\infty} \frac{(-i\theta A)^n}{n!}.
\]
Because \( A^2 = A \), we observe:
\[
A^n = A \quad \text{for all } n \geq 1.
\]
Therefore:
\[
(-i\theta A)^n = (-i\theta)^n A^n = (-i\theta)^n A \quad \text{for } n \geq 1.
\]

We now separate the \( n = 0 \) term and group the rest:
\[
e^{-i \theta A} = I + \sum_{n=1}^{\infty} \frac{(-i\theta)^n}{n!} A.
\]

Factor out \( A \) from the sum:
\[
e^{-i \theta A} = I + \left( \sum_{n=1}^{\infty} \frac{(-i\theta)^n}{n!} \right) A.
\]

This sum is just the exponential function minus the \( n = 0 \) term:
\[
\sum_{n=1}^{\infty} \frac{(-i\theta)^n}{n!} = e^{-i\theta} - 1.
\]

Thus,
\[
e^{-i \theta A} = I + (e^{-i\theta} - 1) A.
\]

\end{document}